\documentclass[english,aps,preprint]{revtex4}
\usepackage[T1]{fontenc}
\usepackage[latin9]{inputenc}
\usepackage{amsmath}
\usepackage{graphicx}
\usepackage{esint}

\makeatletter
\@ifundefined{textcolor}{}
{%
 \definecolor{BLACK}{gray}{0}
 \definecolor{WHITE}{gray}{1}
 \definecolor{RED}{rgb}{1,0,0}
 \definecolor{GREEN}{rgb}{0,1,0}
 \definecolor{BLUE}{rgb}{0,0,1}
 \definecolor{CYAN}{cmyk}{1,0,0,0}
 \definecolor{MAGENTA}{cmyk}{0,1,0,0}
 \definecolor{YELLOW}{cmyk}{0,0,1,0}
}

\makeatother

\usepackage{babel}
\begin{document}

\preprint{This line only printed with preprint option}

\title{The Application of Mutual Energy Theorem in Expansion of Radiation
Field in Spherical Waves}

\author{Shuang-ren Zhao}

\email{Xidian University,Xi'an,China}

\homepage{http://imrecons.com/pu/}

\selectlanguage{english}%

\thanks{It has been received in May of 1985. It has been received and finalized
in June of 1986, it has been published in published on ACTA Electronica
Sinica, Vol. 15, No. 3 May 1987. P. R. of China, P88. Now it is translated
to English.}

\affiliation{Northwest Telecommunication Engineering Institute, Xi\textquoteright an,
China}
\begin{abstract}
``Mutual energy theorem'' and the concept of inner product of electromagnetic
fields are introduced, on which the method of expansion of radiation
field in spherical waves is discussed.
\end{abstract}
\maketitle

\section{introduction}

In recent years the spherical wave expansion method has been widely
applied to the theory and calculation of electromagnetic fields. But
the inner product exist in reference\cite{AWRudge} is defined on
the Banach space\cite{WeixingZheng}. Through redefining the inner
product this article limits the wave expansion method to Hibert space\cite{DCStinson}.
For this reason the mutual energy theorem is introduced.

\section{The definition of the inner product and wave expansions}

The electromagnetic fields in the space can be seen as an element
which can be expressed as,

\begin{equation}
\zeta=\{E,H\}\label{eq:1}
\end{equation}
All the electromagnetic fields in the space compose a set which can
be written as $Q$. In practices that $Q$ is the solution set of
the Maxwell's equations:
\begin{equation}
\{\zeta|\nabla\times E=-j\omega\mu H-K,\nabla\times H=j\omega\epsilon E+J\}\label{eq:2}
\end{equation}
where $K$ is the intensity of the magnetic current, $J$ is the intensity
of current, $\omega$ is the frequency, $\epsilon$,$\mu$ are real
number (also can be complex number, noticed by the translator), which
are permittivity and permeability. $Q$ is a linear space in practical.

Assume there is a volume $V$ in the space, which has the boundary
$\varGamma$ . Assume the field $\zeta$ is produced by $J$ and $K$,
then all this kind of elements compose a subspace $G$ (Assume $\zeta$
belong to the retarded potential, there is no advanced potential,
noticed by the translator). The inner product can be defined on $G$, 

\begin{equation}
(\zeta_{1},\zeta_{2})=\oiintop_{\varGamma}\,(E_{1}\times H_{2}^{*}+E_{2}^{*}\times H_{1})\,\hat{n}\,ds\label{eq:3}
\end{equation}
where, $\zeta_{i}=\{E_{i},H_{i}\}$, $i=1,2$, $\hat{n}$ is the normal
vector direct to the outside of the surface $\varGamma$, The symbol
$*$ expresses the complex conjugate. This inner product satisfies
the inner product laws:

(I) $(\zeta,\zeta)\geq0,\ \zeta=0\ \ \ iff\ (\zeta,\zeta)=0$ 

(II) $(\zeta_{1},\zeta_{2})=(\zeta_{2},\zeta_{1})^{*}$

(III) $(\alpha\zeta_{1}+\beta\zeta_{2},\zeta_{3})=\alpha(\zeta_{1},\zeta_{3})+\beta(\zeta_{2},\zeta_{3})$

Where $iff$ means ``if and only if''; $\alpha$,$\beta$ are any
constants, $\zeta_{1}$, $\zeta_{2}$, $\zeta_{3}$$\in G$.

If the above definition of inner product is widened from $G$ to $Q$,
the law (I) does not satisfy. Hence on the linear space $Q$, the
inner product can only been seen as a generalized definition of the
inner product.

From this definition of the inner product, the definition of the norm
is, 
\begin{equation}
||\zeta||=\sqrt{(\zeta,\zeta)}=(2\,Re\oiintop_{\varGamma}E\times H^{*}\cdot\hat{n}ds)^{\frac{1}{2}}\label{eq:4}
\end{equation}
In the above formula, $Re$ means ``taking real part''. 

Assume that $\{\zeta_{\lambda}\}$ is a complete set on $G$. Here
$\lambda\in\Lambda$, $\Lambda$ is a index set, which satisfies the
normalized orthogonal condition,

\begin{equation}
(\zeta_{\mu,}\zeta_{\nu})=\delta_{\mu\nu}\ \ \ \ \ (\mu,\nu\in\Lambda)\label{eq:5}
\end{equation}
where, $\delta_{\mu\nu}$ is Kronecker operator. For any $\zeta\in G$
there is 
\begin{equation}
\zeta=\sum_{\lambda\in\Lambda}c_{\lambda}\zeta_{\lambda}\label{eq:6}
\end{equation}
where the expansion coefficient can be found as $c_{\lambda}=(\zeta,\zeta_{\lambda})$.

For spherical expansion, $\{\zeta_{\lambda}\}$ is chosen as spherical
wave function, in this case $\lambda$ means $nm$, $\zeta_{\lambda}$
have two forms $\xi_{nm},\eta_{nm}$, where 
\begin{equation}
\xi_{nm}=\{M_{nm},\frac{j}{\eta}N_{nm}\}\label{eq:7}
\end{equation}
\begin{equation}
\eta_{nm}=\{N_{nm},\frac{j}{\eta}M_{nm}\}\label{eq:8}
\end{equation}
where the constant factor $\frac{j}{\eta}$ make the above formula
so that if the first item inside the brace $\{\}$ is electric field,
the second will correspond to the magnetic field. $M_{nm}=\nabla\times[g_{n}\vec{r}h_{n}(kr)Y_{n}^{m}(\theta,\varphi)$,
$N_{nm}=\frac{1}{K}\nabla\times M_{nm}$. $h_{n}$ is $n$ level class
II spherical Hankel function. $k=\omega\sqrt{\epsilon\mu}$, $\eta=(\mu/\epsilon)^{\frac{1}{2}}$,
$g_{n}$ is a normalized constant which will be found later. $r,\theta$
and $\varphi$ are spherical coordinates. The corresponding unit vector
are $\hat{r}$, $\hat{\theta}$, $\hat{\varphi}$, but $\vec{r}=r\hat{r}$,
$Y_{n}^{m}$are $nm$ level spherical function, 
\begin{equation}
Y_{n}^{m}(\theta,\varphi)=[(2n+1)\frac{(n-m)!}{(n+m)!}]^{\frac{1}{2}}P_{n}^{m}(\cos\theta)e^{jm\varphi}\label{eq:9}
\end{equation}
where, $P_{n}^{m}$ is associated Legendre function, $n=0,1\cdots\cdots,$
$m=\pm0,\pm1\cdots\cdots\pm n$.

If the normalized orthogonal are established for the spherical wave,
Eq.(\ref{eq:6}) can be rewritten as,

\begin{equation}
\left\{ \begin{array}{c}
\zeta=\sum_{nm}(a_{nm}\xi_{nm}+b_{nm}\eta_{nm}\\
a_{nm}=(\zeta,\xi_{nm}),\ \ b_{nm}=(\zeta,\eta_{nm})
\end{array}\right\} \label{eq:10}
\end{equation}

\section{The orthogonalization and normalization of the spherical wave}

It can be seen from Eq.(\ref{eq:4}) that, if the sources of $\zeta$
is inside the surface $\varGamma$, i.e. $\zeta\in G$, the norm $||\zeta||$
is only the power flow to the outside of $\varGamma$. If the media
are lossless, then the power of $\zeta$ is not changed to different
surface $\varGamma$. Considering a surface $\varGamma$ with arbitrary
radio, we can obtain from the calculation{[}3{]} that,

\begin{equation}
||\zeta_{\lambda}||^{2}=||\xi_{nm}||^{2}=||\eta_{nm}||^{2}=\frac{8\pi}{\eta k^{2}}n(n+1)g_{n}\label{eq:11}
\end{equation}
Considering the normalization condition $||\zeta_{\lambda}||=1$,
the normalization constant can be obtained

\begin{equation}
g_{n}=k\sqrt{\frac{\eta}{8\pi(n+1)}}\label{eq:12}
\end{equation}

For convenience, the spherical wave is divided to 

\begin{equation}
\begin{cases}
\xi_{nm}=\xi_{nm}^{(1)}+\xi_{nm}^{(2)}\\
\eta_{nm}=\eta_{nm}^{(1)}+\eta_{nm}^{(2)}
\end{cases}\label{eq:13}
\end{equation}

\begin{equation}
\begin{cases}
\xi_{nm}^{(i)}=\{M_{nm}^{(i)},\frac{j}{\eta}N_{nm}^{(i)}\}\\
\eta_{nm}^{(i)}=\{N_{nm}^{(i)},\frac{j}{\eta}M_{nm}^{(i)}\}
\end{cases}(i=1,2)\label{eq:14}
\end{equation}

(Notice: I have corrected the print error in the above formula, in
the original publication, $\xi_{nm}^{(i)}$ is written as $\xi_{1}^{(i)}$
and $\eta_{nm}^{(i)}$ is written as $\eta^{(i)}$, noticed by the
translator.) 
\begin{equation}
\begin{cases}
M_{nm}^{(1)}=\nabla\times[g_{n}\overrightarrow{r}j_{n}(kr)Y_{n}^{m}(\theta,\varphi)\\
M_{nm}^{(2)}=\nabla\times[-jg_{n}\overrightarrow{r}\,n_{n}(kr)Y_{n}^{m}(\theta,\varphi)\\
N_{nm}^{(i)}=\frac{1}{k}\nabla\times M_{nm}^{(i)} & (i=1,2)
\end{cases}\label{eq:15}
\end{equation}

Actually the above is that the second class spherical Hankel function
$h_{n}(kr)=j_{n}(kr)-jn_{n}(kr)$ in the wave function is divided
to two items. $j_{n}$,$n_{n}$ are first class and second class spherical
Bessel functions.

According to Eq.(12) and reference\cite{AWRudge} it can be proven
that the spherical wave are orthogonal 

\begin{equation}
(\xi_{nm},\xi_{ts})=\delta_{nt}\delta_{ms}\label{eq:16}
\end{equation}
\begin{equation}
(\xi_{nm},\xi_{ts}^{(i)})=(\xi_{nm},\xi_{ts}^{(i)})=\frac{1}{2}\delta_{nt}\delta_{ms}\ \ \ \ \ \ (i=1,2)\label{eq:17}
\end{equation}
\begin{equation}
(\xi_{nm}^{(i)},\eta_{ts}^{(j)})=0\label{eq:18}
\end{equation}
In the above 3 formula $n$, $m$, $t$, $s$ are integer. in Eq.(\ref{eq:18})
$i$,$j$ are $1,2$ or no superscript.

\section{The mutual energy theorem and the modified mutual energy theorem}

A theorem is introduced, which is referred as the mutual energy theorem
\[
\oiint_{\Gamma}(E_{1}\times H_{2}^{*}+E_{2}^{*}\times H_{1})\,\hat{n}dS
\]
\begin{equation}
=-\intop_{V}(J_{1}\cdot E_{2}^{*}+K_{1}\cdot H_{2}^{*}+J_{2}^{*}\cdot E_{1}+K_{2}^{*}\cdot H_{1})\,dV\label{eq:19}
\end{equation}
 In the above formula $\varGamma$ is the boundary of volume $V$.
$\hat{n}$, normal vector directed to the outside of surface $\varGamma$.
$J_{1},K_{1}$and $J_{2},K_{2}$ are the sources of field $E_{1},H_{1}$
and $E_{1},H_{2}$. This theorem is established in the lossless media,
that is if the media satisfies that,

\begin{equation}
\overleftrightarrow{\epsilon}=(\overleftrightarrow{\epsilon})^{\dagger}\ \ \ \ \ \ \ \overleftrightarrow{\mu}=(\overleftrightarrow{\mu}){}^{\dagger}\label{eq:20}
\end{equation}
In the above formula, the superscript $\dagger$ is matrix transpose
and complex conjugate. This theorem is proven in the following.

Recent year in the four wave frequency mixing theory\cite{PeiXuanYie},
the concept of conjugate wave has been applied. This concept can be
summarized as conjugate transform.

If $E,H$ satisfy the Maxwell's equations

\begin{equation}
\nabla\times E=-j\omega\overleftrightarrow{\mu}H-K\ \ \ \ \ \ \ \nabla\times H=+j\omega\overleftrightarrow{\epsilon}E+J\label{eq:21}
\end{equation}
Then take the transform $E=E^{b*}$, $H=-H^{b*}$, $K=K^{b*},$ $J=-J^{b*}$,
$\overleftrightarrow{\epsilon}=(\overleftrightarrow{\epsilon})^{b*}$,
$\overleftrightarrow{\mu}=(\overleftrightarrow{\mu})^{b*}$, $E^{b},H^{b}$
still satisfy the Maxwell's Equations Eq.(\ref{eq:21}). Hence this
transform is referred as conjugate transform. 

To the variable with subscript $2$ in the reversible theorem in reference\cite{JAKong}
make the conjugate transform the modified mutual energy theorem can
be obtained. 
\[
\oiint_{\Gamma}(E_{1}\times H_{2}^{d*}+E_{2}^{d*}\times H_{1})\,\hat{n}dS
\]
\begin{equation}
=-\intop_{V}(J_{1}\cdot E_{2}^{d*}+K_{1}\cdot H_{2}^{d*}+J_{2}^{d*}\cdot E_{1}+K_{2}^{d*}\cdot H_{1})\,dV\label{eq:22}
\end{equation}
In the above formula $E_{2}^{d},H_{2}^{d}$ satisfy Maxwell's equation
with the associate media $(\overleftrightarrow{\epsilon})^{d}$, $(\overleftrightarrow{\epsilon})^{d}$.
Here 
\[
(\overleftrightarrow{\epsilon})^{d}=(\overleftrightarrow{\epsilon})^{\dagger}
\]
\[
(\overleftrightarrow{\mu})^{d}=(\overleftrightarrow{\mu})^{\dagger}
\]

If the media is lossless, i.e. Eq(\ref{eq:20}) is established, then
the associated media is same to the original media, ($(\overleftrightarrow{\epsilon})^{d}=\overleftrightarrow{\epsilon},\ (\overleftrightarrow{\mu})^{d}=\overleftrightarrow{\mu}$).
In this time the modified mutual energy theorem become the mutual
energy theorem Eq.(\ref{eq:19}).

In the following assume the media are isotropic. Hence the tensors
permittivity $\overleftrightarrow{\epsilon}$ and permeability $\overleftrightarrow{\mu}$
become constant permittivity $\epsilon$ and permeability $\mu$.

\section{The formula for the coefficient of the spherical wave expansion}

\subsection{Obtaining the coefficients of the spherical expansion knowing the
current distribution}

Assume the source of radiation is the current distribution $J$, $J$
is inside the spherical surface $\varGamma$. Apply the mutual energy
theorem on $\varGamma$, we can obtain,

\[
c_{\lambda}=(\zeta,\zeta_{\lambda})=\oiintop_{\varGamma}\ (E\times H_{\lambda}^{*}+E_{\lambda}^{*}\times H)\,\hat{n}ds
\]

\begin{equation}
-\intop_{V}E_{\lambda}^{*}\cdot J\,dv-\oiintop_{\varGamma_{\delta}}\,(E\cdot J_{\lambda}^{*}+H\cdot K_{\lambda}^{*})\,ds\label{eq:23}
\end{equation}
In the above formula $\zeta_{\lambda}=\{E,H\}$, and $E_{\lambda}=M_{nm}$
or $N_{nm}$, $J_{\lambda}$and $K_{\lambda}$are effective current
and magnetic current sources of the filed $\zeta_{\lambda}$. They
are distributed on the spherical surface $\varGamma_{\delta}$ which
has radio $\delta\rightarrow0$ and its center at the oringin $O$.
(Notice, according to the mutual energy theorem formula Eq.(\ref{eq:19})
the integral $\oiintop_{\varGamma_{\delta}}\,(E\cdot J_{\lambda}^{*}+H\cdot K_{\lambda}^{*})\,ds$
is an integral on a volume $V_{\delta}$, the boundary surface of
the volume $V_{\delta}$ is $\varGamma_{\delta}$; in the limit situation,
the volume integral changed to surface integral. Noticed by the translator.)

For the special situation that the field $\zeta=\{E,H\}$ is just
zero at the epsilon neighborhood to origin, the last item of the formula
Eq.(\ref{eq:23}) vanishes. This situation same as the origin is chosen
inside the antenna, the antenna can be seen as ideal electric conductor.
In this situation there are

\begin{equation}
\begin{cases}
c_{\lambda}=-\intop_{V}E_{\lambda}^{*}J\,dv\\
a_{nm}=-\intop_{V}M_{nm}^{*}J\,dv\\
b_{nm}=-\intop_{V}N_{nm}^{*}J\,dv
\end{cases}\label{eq:24}
\end{equation}

In case in the epsilon neighborhood the field is not zero, in the
Eq.(\ref{eq:23}) can have integral divergence. In order to overcome
this difficulty, assume that the effective source $J_{\lambda}$,
$K_{\lambda}$ produce the filed outside of the surface $\varGamma_{\lambda}$
is $\zeta_{\lambda}$, and inside the $\varGamma_{\lambda}$ is $\zeta_{\lambda}^{(1)}$.
$\zeta_{\lambda}^{(1)}=\xi_{nm}^{(1)}$ or $\eta_{nm}^{(1)}$ which
is given through Eq.(\ref{eq:14}).

\begin{figure}
\includegraphics[scale=0.1]{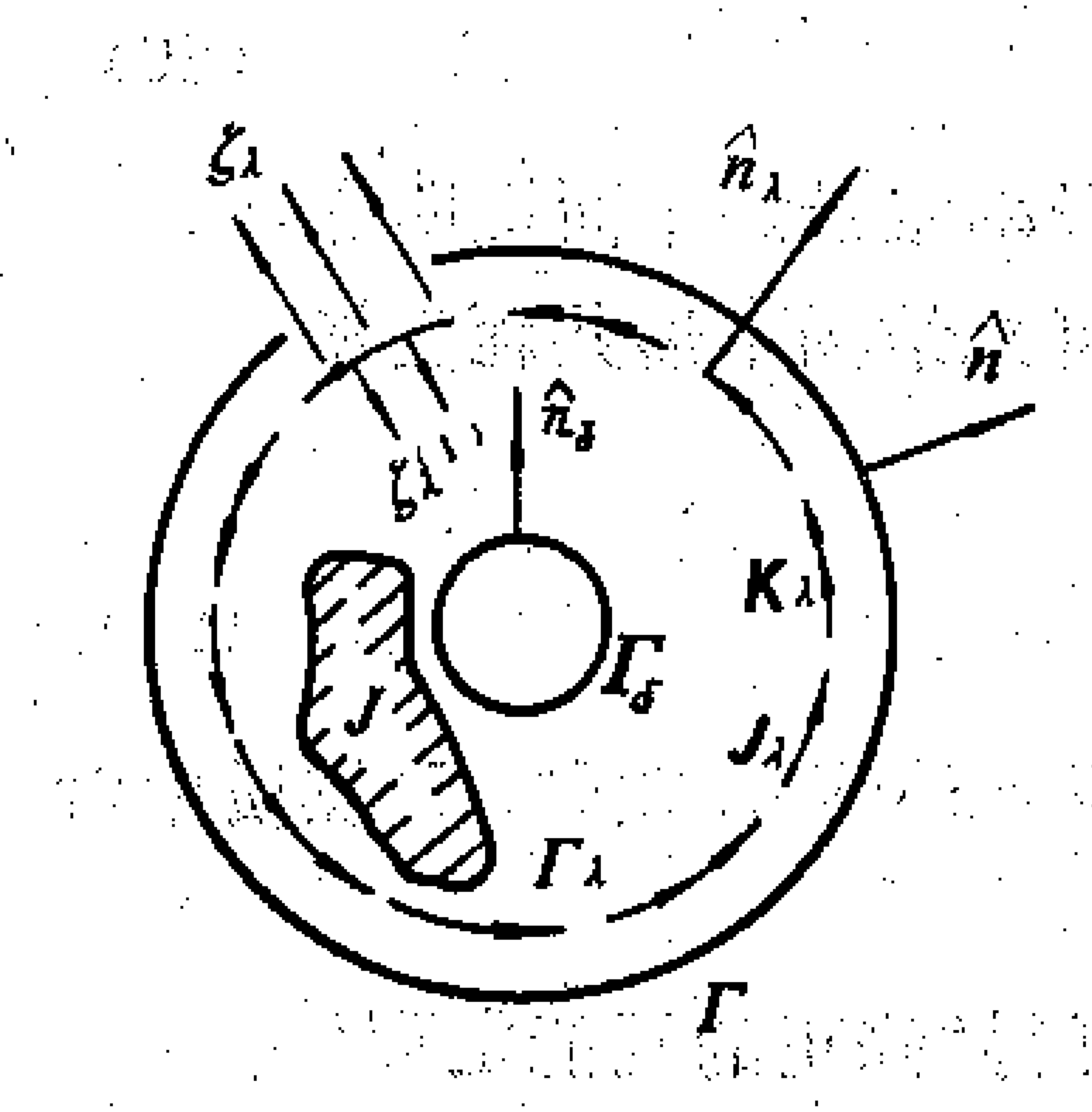}\label{1}

\caption{Effective Source $J_{\lambda},$$K_{\lambda}$, radiation source $J$,
and the surface $\varGamma$,$\varGamma_{\delta}$,$\varGamma_{\lambda}$.}
\end{figure}

According to effective principle that the effective source can be
written as
\begin{equation}
\begin{cases}
J_{\lambda}=\hat{n}_{\lambda}\times(H_{\lambda}-H_{\lambda}^{(1)})=\hat{n}_{\lambda}\times H^{(2)}\\
K_{\lambda}=-\hat{n}_{\lambda}\times(E_{\lambda}-E_{\lambda}^{(1)})=-\hat{n}_{\lambda}\times E_{\lambda}^{(2)}
\end{cases}\label{eq:25}
\end{equation}
In the above formula, $\hat{n}_{\lambda}$ is outside direction normal
vector of the spherical surface $\varGamma_{\lambda}$. $\zeta_{\lambda}^{(2)}=(E_{\lambda}^{(2)},H_{\lambda}^{(2)})$
can be calculated through Eq.(\ref{eq:14}). The source distribution
can be seen in Figure \ref{1}. 

This way the effective source $J_{\lambda}$, $K_{\lambda}$can create
transmission wave at outside the spherical surface and produce standing
wave inside the shperical surface. Apply mutual energy theorem on
the spherical surface $\varGamma$ can obtain,

\[
c_{\lambda}=(\zeta,\zeta_{\lambda})=\oiintop_{\varGamma}\,(E\times H_{\lambda}^{*}+E_{\lambda}^{*}\times H)\cdot\hat{n}ds
\]
\begin{equation}
=-\intop_{V}E_{\lambda}^{(1)}\cdot Jdv-\oiintop_{\varGamma_{\lambda}}\,(E\cdot J_{\lambda}^{*}+H\cdot K_{\lambda}^{*})\,ds\label{eq:26}
\end{equation}
(Notice, $E_{\lambda}^{(1)}$, $H_{\lambda}^{(1)}$ are transmission
waves, $E_{\lambda}^{(2)}$, $H_{\lambda}^{(2)}$ are stand vaves,
Noticed by the translator)

Considering Eq.(\ref{eq:25}), the last item in the above formula
is,
\begin{equation}
-\oiintop_{\varGamma_{\lambda}}(E^{*}\cdot J_{\lambda}+H^{*}\cdot K_{\lambda})\,ds=(\zeta,\zeta_{\lambda}^{(2)})\label{eq:27}
\end{equation}
And from orthogonal condition Eq.(\ref{eq:17}) we can obtain,

\begin{equation}
(\zeta,\zeta_{\lambda}^{(2)})=(\sum_{\lambda\in\Lambda}c_{\lambda}\zeta_{\lambda},\zeta_{\lambda}^{(2)})=\frac{1}{2}c_{\lambda}\label{eq:28}
\end{equation}

Substituting Eq(\ref{eq:27}), Eq.(\ref{eq:28}) to Eq(\ref{eq:26})
and after arrangement we obtain,

\begin{equation}
\begin{cases}
c_{\lambda}=-2\intop_{V}E_{\lambda}^{(1)*}\cdot J\,dv\\
a_{nm}=-2\intop_{V}M_{nm}^{(1)*}\cdot J\,dv\\
b_{nm}=-2\intop_{V}N_{nm}^{(1)*}\cdot J\,dv
\end{cases}\label{eq:29}
\end{equation}
This result is same as reference\cite{CHPapas}.

\subsection{the coefficients of the spherical expansion knowing the tangential
component of the electromagnetic fields }

Assume there is radiation source current intensity $J$ inside the
sphere $\varGamma$. This source produce the field $\zeta=\{E,H\}$.
We have measured the tangent filed on the spherical surface $\varGamma$.
The wave can be expanded as spherical wave, the spherical wave coefficients
are,

\begin{equation}
\begin{cases}
a_{nm}=\oiintop_{\varGamma}[-\frac{j}{\eta}(E_{\theta}N_{nm\varphi}^{*}-E_{\varphi}N_{nm\theta}^{*})-H_{\theta}M_{nm\varphi}^{*}+H_{\varphi}M_{nm\theta}^{*}]ds\\
b_{nm}=\oiintop_{\varGamma}[-\frac{j}{\eta}(E_{\theta}M_{nm\varphi}^{*}-E_{\varphi}M_{nm\theta}^{*})-H_{\theta}N_{nm\varphi}^{*}+H{}_{\varphi}N_{nm\theta}^{*}]ds
\end{cases}\label{eq:30}
\end{equation}
In the above formula, $X_{\theta}$$X_{\varphi}$ are tangential component
of $X$ in $\theta$ and $\varphi$ direction.

\subsection{The method to find the coefficients of the spherical wave expansion
for the radiation sources when there exist the interference sources}

Assume there radiation source $J$ and the interference source $J_{g}$.
The corresponding field is $\zeta_{g}=\{E_{g},H_{g}\}$. Chosen the
sphere $\varGamma$ so that $J$ is inside $\varGamma$, $J_{g}$
is outside the $\varGamma$. Let the sum of $J$ and $J_{g}$ is $J'$,
the corresponding field is $\zeta'=\{E',H'\}$. In the boundary $\varGamma$
measured that the tangential field of $E'$ and $H'$.

\begin{figure}
\includegraphics{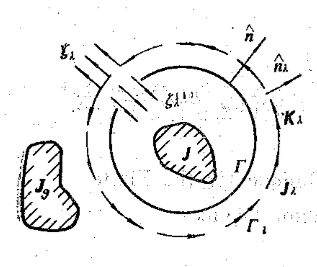}\label{2}

\caption{The effective source $J_{\lambda},$$K_{\lambda},$ radiation source
$J$, and the interference source $J_{g}$}
\end{figure}

$\varGamma_{\lambda}$ is the source of spherical wave $\zeta_{\lambda}$.
$\varGamma_{\lambda}$ is chosen at the outside of the spherical surface
$\varGamma$ but let the interference source be at the outside of
$\varGamma_{\lambda}$. In this situation the effective source still
can be given by Eq.(\ref{eq:25}). This way $J_{\lambda}$ , $K_{\lambda}$will
produce the field $\zeta_{\lambda}^{(1)}$ inside the spherical surface
$\varGamma_{\lambda}$ and produce the field $\zeta_{\lambda}$ outside
the spherical surface $\varGamma_{\lambda}$. All sources and fields
and surfaces can be seen in Figure \ref{2}.

Taking the inner product at the surface $\varGamma$ and considering
the orthogonal formula Eq.(\ref{eq:17}), we can obtain, 

\begin{equation}
(\zeta,\zeta_{\lambda}^{(1)})=(\sum_{\lambda\in\Lambda}c_{\lambda}\zeta_{\lambda,}\zeta_{\lambda}^{(1)})=\frac{1}{2}c_{\lambda}\label{eq:31}
\end{equation}
Because the radiation field $\zeta=\zeta'-\zeta_{g}$, there is,

\begin{equation}
(\zeta,\zeta_{\lambda}^{(1)})=(\zeta',\zeta_{\lambda}^{(1)})-(\zeta_{g},\zeta_{\lambda}^{(1)})\label{eq:32}
\end{equation}

Considering inside the spherical surface $\varGamma$, $J_{g}$ which
is the the source of field $\zeta_{g}$ and effective source $J_{\lambda}$and
$K_{\lambda}$are all vanishes, according to the mutual energy theorem
Eq.(\ref{eq:19}) the last item of Eq.(\ref{eq:32}) vanishes:

\begin{equation}
(\zeta_{g},\zeta_{\lambda}^{(1)})=\oiintop_{\varGamma}\,(E_{g}\times H_{\lambda}^{(1)*}+E_{\lambda}^{(1)*}\times H_{g})\cdot\hat{n}ds=0\label{eq:33}
\end{equation}
Hence, there is
\begin{equation}
c_{\lambda}=2(\zeta',\zeta_{\lambda}^{(1)})\label{eq:34}
\end{equation}

\begin{equation}
\begin{cases}
a_{nm}=2\oiintop_{\varGamma}[-\frac{j}{\eta}(-E'_{\theta}N_{nm\varphi}^{(1)*}-E'_{\varphi}N_{nm\theta}^{(1)*})-H_{\theta}'M_{nm\varphi}^{(1)*}+H'_{\varphi}M_{nm\theta}^{(1)*}]ds\\
b_{nm}=2\oiintop_{\varGamma}[-\frac{j}{\eta}(-E'_{\theta}M_{nm\varphi}^{(1)*}-E'_{\varphi}M_{nm\theta}^{(1)*})-H_{\theta}'N_{nm\varphi}^{(1)*}+H'_{\varphi}N_{nm\theta}^{(1)*}]ds
\end{cases}\label{eq:35}
\end{equation}

Comparing with Eq.(\ref{eq:30}), the advantage of Eq.(\ref{eq:35})
is that there is no influence of the obtained coefficient of the spherical
wave expansion with the interference sources. 

For the problem of measurement of scattering field, this result can
be applied. The incident field and the scattering field can be seen
as interference field and the radiation field. This way the calculation
of the expansion coefficient of the spherical wave for the scattering
field can directly use the formula Eq.(\ref{eq:35}) . Hence if the
superimposed field of the incident field and the scattering field
have been measured, the scattering field can be calculate at the outside
of the spherical surface $\Gamma$.

\section{Conclusion}

When making spherical expansion, handiness apply mutual energy theorem
can simplify the derivation process and make the physical meaning
clearer. All the more so in the case the influence of the interference
source must be considered.
\begin{acknowledgments}
Thanks the help of the teacher: Liu Pencheng, Deming Fu, Changhong
Liang, Deshuan Yang.\end{acknowledgments}

\end{document}